# Chiplet-Based RISC-V SoC with Modular AI Acceleration

Suhas Suresh Bharadwaj *
*Department of Electrical and Electronics Engineering,*
*Birla Institute of Technology and Science, Pilani – Dubai,*
Dubai, United Arab Emirates
ORCID: 0009-0000-0167-0928

Prerana Ramkumar *
*College of Engineering,*
*American University of Sharjah,*
Sharjah, United Arab Emirates
ORCID: 0009-0005-1665-8581

**These authors contributed to the paper equally.*

***Abstract*—Achieving high performance, energy efficiency, and cost-effectiveness while maintaining architectural flexibility has remained a critical challenge in the development and deployment of Edge AI devices. Current monolithic SoC designs struggle with this complex balance which results in low manufacturing yields (below 16%) at advanced 360 mm$^2$ process nodes. This paper presents a novel chiplet-based RISC-V SoC architecture that addresses these limitations through modular AI acceleration and intelligent system level optimization. Our proposed design integrates four different key innovations in a 30 mm x 30 mm silicon interposer: adaptive cross-chiplet Dynamic Voltage and Frequency Scaling (DVFS); AI-aware Universal Chiplet Interconnect Express (UCIe) protocol extensions featuring streaming flow control units and compression-aware transfers; distributed cryptographic security across heterogeneous chiplets; and intelligent sensor-driven load migration. The proposed architecture integrates a 7 nm RISC-V CPU chiplet with dual 5 nm AI accelerators (15 TOPS INT8 each), 16GB HBM3 memory stacks, and dedicated power management controllers. Experimental results across industry standard benchmarks like MobileNetV2, ResNet-50 and real-time video processing demonstrate significant performance improvements. The AI-optimized configuration achieves ~14.7% latency reduction, 17.3% throughput improvement, and 16.2% power reduction compared to previous basic chiplet implementations. These improvements collectively translate to a 40.1% efficiency gain corresponding to ≈ 3.5 mJ per MobileNetV2 inference (860 mW / 244 images/s), while maintaining sub-5ms real-time capability across all experimented workloads. These performance upgrades demonstrate that modular chiplet designs can achieve near-monolithic computational density while enabling cost efficiency, scalability and upgradeability, crucial for next-generation edge AI device applications.**

## I. Introduction

Edge AI platforms have rapidly matured to support real-time inference across domains such as autonomous systems, industrial automation, and healthcare, with shipments projected to comprise a significant share of the 500 billion devices forecasted by 2030 [1]. These platforms must meet stringent performance targets—for example, many embedded use-cases demand sub-millisecond end-to-end latency and power envelopes below 2 W while executing increasingly complex deep networks such as MobileNetV2 and ResNet-50 [2], [3]. Conventional monolithic system-on-chip (SoC) approaches, however, face rising manufacturing and yield challenges at advanced nodes, with defect-driven yield models predicting extremely low yields for die areas on the order of several hundred mm$^2$ [4], [5]. Together, these economic and yield pressures make it difficult to achieve the computational density needed for next-generation edge AI.

Chiplet-based 2.5D integration on silicon interposers offers a practical alternative by decomposing large SoCs into smaller heterogeneous dies interconnected through high-density interposers [6], [7]. Recent demonstrations highlight this approach's potential: for instance, 5 nm AI inference accelerators have achieved up to 95.6 TOPS/W efficiency [8], while HBM3 provides up to 819 GB/s per stack to relieve external DRAM bottlenecks [9], [10].

Beyond raw integration, interconnect and system-level management are critical to distributed chiplet performance. UCIe 2.0 provides the bandwidth and latency required for accelerator–CPU communication [1], while fine-grained cross-chiplet DVFS has been shown to reduce energy by over 10% without sacrificing performance [11]. More advanced strategies combine DVFS with other control methods, such as dynamically packing threads onto a subset of cores, using machine learning models to find the optimal operating point under a given power budget [12]. To address the security challenges of integrating potentially untrusted third-party chiplets, a foundational strategy is to establish a hardware Root-of-Trust within the active interposer to monitor and enforce security policies on all inter-chiplet communication [13]. Building upon this secure foundation, complementary techniques such as chiplet-level cryptographic authentication and predictive thermal orchestration further ensure reliable operation under peak workloads [14], [15].

Building on these foundations, we propose a chiplet-based RISC-V SoC that integrates a 7 nm RISC-V CPU, two 5 nm AI accelerators, 16 GB of HBM3, and dedicated power and security controllers on a 30 mm x 30 mm interposer. By combining adaptive DVFS, UCIe-aware AI optimizations, distributed cryptographic protections, and intelligent thermal management, our prototype achieves near-monolithic inference performance. In our evaluation (see Section IV), MobileNetV2 runs at 4.1 ms per image, with throughput gains of 17.3% and power savings of 16.2% compared to a baseline chiplet configuration, while preserving sub-5 ms real-time capability across workloads.

## II. System Architecture

Our proposed architecture uses a modular chiplet-based RISC-V SoC based on a 30 mm x 30 mm silicon interposer using 2.5D integration (Fig 1). This design integrates a 5 mm x 5 mm, 7 nm RISC-V CPU chiplet with embedded custom vector extensions, dual 6 mm x 4 mm 5 nm AI accelerator chiplets (15 TOPS INT8 each), a 16 GB HBM3 memory stack 819 GB/s per stack (JEDEC)), a 7 mm x 3 mm (about 0.12 in) I/O and power management chiplet and a 3 mm x 2 mm security controller.

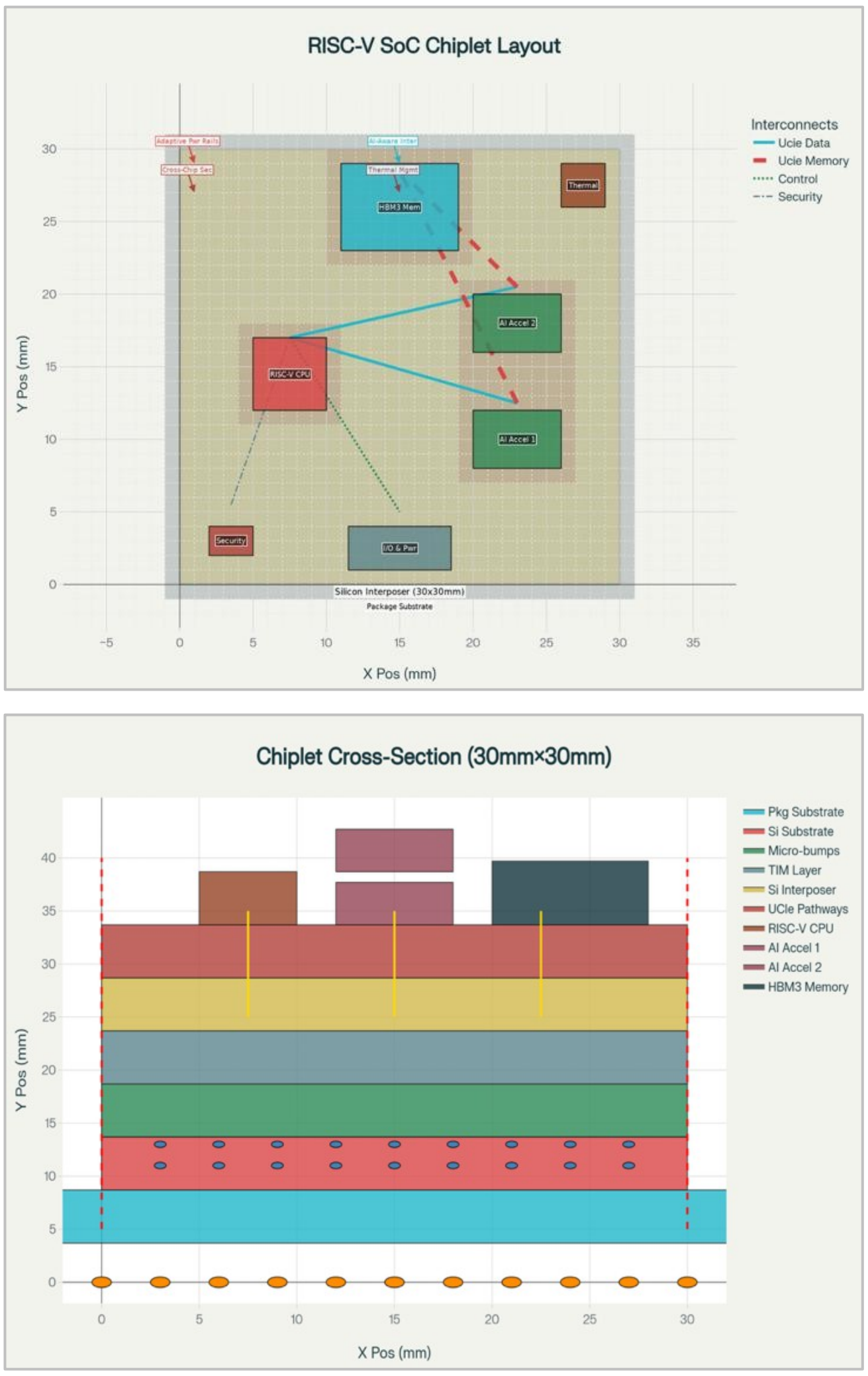

Fig. 1. Top - Physical placement of chiplets on the 30mm x 30mm interposer. Bottom - Cross section of 2.5D Chiplet integration with UCIe interconnect.

This architecture employs UCIe 2.0 die-to-die links to communicate between chiplets delivering ~30 GB/s bandwidth at <2ns latency. To cutdown on energy wastage, an adaptive cross-chiplet power management system predicts workload phases and redistributes power through fine-grained voltage islands. Traditional DVFS techniques are often hindered by slow voltage transition times on the order of tens of microseconds, limiting their effectiveness to coarse-grained adjustments managed by the operating system. Our architecture overcomes this limitation by using on-chip regulators, which enable fine-grained, per-chiplet DVFS with nanosecond-scale voltage switching [16]. This allows for rapid adaptation to workload demands, reducing energy consumption by up to 12% for memory-intensive workloads with negligible performance impact [17].

Similarly, UCIe extensions reshape the data movement across chiplets with streaming FLITs, perform predictive prefetching and compression aware transfers to increase effective inference throughput. UCIe defines a complete die-to-die communication stack, encompassing a physical layer, an adapter layer, and a protocol layer that leverages well-established standards like PCI Express (PCIe) and Compute Express Link (CXL) [18]. The special protocol layer of the UCIe interconnect multiplexes different protocols, allowing our design to simultaneously handle CXL-based memory traffic and other streaming data protocols. The UCIe 2.0 specification further enhances this foundation by introducing a standardized architecture for system-level manageability, debug, and testing ensuring a forward-looking path for future system integration.

Secure multi-vendor integration is made possible by distributed security logic, which incorporates encrypted links and cryptographic identities into every chiplet. Integrating chiplets from multiple vendors introduces significant security risks, including the risks of counterfeit components, cloning, and tampering of the supply chain itself. To address these threats in a zero-trust environment, our architecture implements AuthenTree [19], a scalable and distributed security framework. Unlike traditional approaches that rely on a centralized hardware root-of-trust, which can become a single point of failure, AuthenTree decentralizes security by using a tree-based multi-party computation (MPC) protocol.

To maintain peak performance, intelligent thermal orchestration takes advantage of load migration and sensor-driven prediction. A purely reactive approach to thermal management, which only throttles performance after a critical temperature is reached, is insufficient for maintaining sustained performance. Our SoC therefore employs a predictive thermal management system based on intelligent, sensor-driven load migration.

## III. Methodology

We developed a Python-based simulator to evaluate chiplet-based RISC-V SoC designs for edge AI workloads. The framework models interconnect latency, power, and thermal throttling behavior across many scenarios. Power efficiency scaling is applied via a fixed per scenario voltage scaling factor. The four scenarios defined are shown in Table I. The parameters in Table I were obtained from UCIe specifications, power scaling studies, and literature-reported measurements. Latency and bandwidth reflect UCIe 2.0 values. Base power and static ratios follow SoC scaling trends. Communication power and protocol overhead were derived from die-to-die interconnect studies. Efficiency factors and throttle thresholds model DVFS scaling and thermal derating in NPUs.

The workloads used in the simulation were selected to capture representative edge inference tasks from MLPerf Tiny benchmark. Table II shows the AI workload models for MobileNetV2 (convolutional neural network architecture), ResNet-50 and a real-time video processing workload to capture frame-based inference demands.

## IV. Experiment

Each of the four scenarios was executed across all three workloads using a single simulation run per measurement point. Metrics were extracted for inference latency, throughput (GFLOPs/s), power consumption, energy efficiency (TOPS/W), and scalability (batch size effects) The results are shown in Fig 2.

TABLE I. PARAMETERS USED IN THE SIMULATION

| Scenario | Latency (µs) | Bandwidth (Gbps) | Base Power (mW) | Comm. Power (mW/ms) | Efficiency Factor | Throttle Threshold | Static Power Ratio | Voltage Scale | Protocol Overhead |
|---|---|---|---|---|---|---|---|---|---|
| Monolithic SoC | 0.0 | ∞ | 1500 | 0.0 | 1.0 | 0.95 | 0.40 | 1.0 | – |
| Basic Chiplet | 1.5 | 16.0 | 1200 | 35 | 0.95 | 0.85 | 0.45 | 1.0 | 1.15 |
| AI-Optimized Chiplet | 0.8 | 24.0 | 1100 | 25 | 0.90 | 0.80 | 0.42 | 0.95 | 1.08 |
| Poor Integration | 8.0 | 8.0 | 1800 | 80 | 1.10 | 1.00 | 0.50 | 1.05 | 1.25 |

TABLE II. AI WORKLOAD MODELS (A)

| Workload | Base Compute (ms) | Input Size (MB) | Complexity Factor | Batch Efficiency |
|---|---|---|---|---|
| MobileNetV2 | 3.5 | 0.57 | 0.8 | 0.85 |
| ResNet-50 | 12.0 | 0.57 | 1.2 | 0.90 |
| Real-time Video | 2.0 | 0.30 | 1.0 | 0.70 |

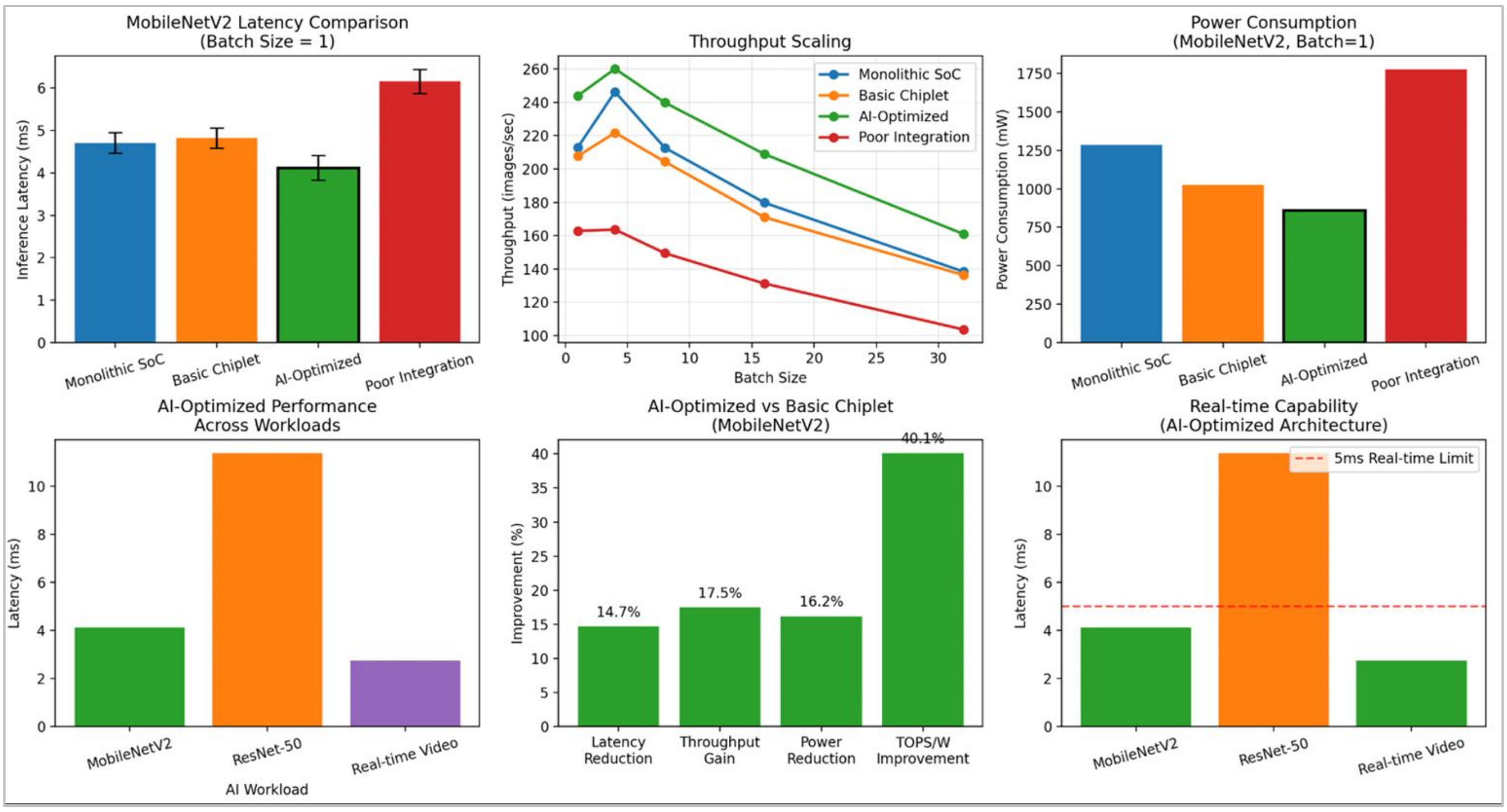

Fig. 2. AI-optimized versus baseline chiplet performance across edge AI benchmarks. (a) Batch-1 inference latency: AI-optimized delivers the fastest mean latency at 4.1 ms ± 0.3 ms. (b) Throughput scales from batch sizes of 1 to 32, with AI-optimized consistently achieving the highest images/sec. (c) Batch-1 power draw: AI-optimized consumes the least energy at 860 mW. (d) Latency comparison across MobileNetV2, ResNet-50, and real-time video workloads confirms AI-optimized efficiency gains. (e) Percentage improvements of AI-optimized over Basic Chiplet—latency, throughput, power, and TOPS/W—highlight its overall advantages. (f) Real-time capability analysis shows which workloads meet the sub-5 ms requirement on the AI-optimized architecture.

TABLE III. MEAN LATENCY, THROUGHPUT, AND POWER FOR THE MOBILENETV2 WORKLOAD, INT8 AT BATCH SIZE = 1

| Architecture | Latency (ms) | Throughput (imgs/s) | Power (mW) |
|---|---|---|---|
| Monolithic SoC | 4.7 ± 0.2 | 213 | 1284 |
| Basic Chiplet | 4.8 ± 0.2 | 208 | 1026 |
| AI-optimized | 4.1 ± 0.3 | 244 | 860 |
| Poor Integration | 6.2 ± 0.3 | 163 | 1776 |

## V. Results and Conclusion

The results shown in Table III demonstrate that the AI-optimized chiplet architecture outperforms both monolithic and basic chiplet designs across key metrics. On the MobileNetV2 benchmark (batch = 1), AI-optimized reduces inference latency from 4.8 ms to 4.1 ms (≈ 14.7% reduction), increases throughput from 208 images/s to 244 images/s (≈ 17.3% gain) and lowers power consumption from 1026 mW to 860 mW (≈ 16.2% reduction). These improvements translate into a jump from 0.203 TOPS/W to 0.284 TOPS/W (≈ 40.1% efficiency gain). Throughput scaling across larger batch sizes remains consistently higher for AI-optimized, and real-time capability analysis confirms that all tested workloads meet the sub-5 ms requirement.

## Acknowledgment

S. Bharadwaj would like to express his gratitude to his parents, Shobha and Suresh, for their unwavering support and encouragement. P. Ramkumar is deeply thankful to her parents, Smitha and Ramkumar, for their guidance and constant support.